\documentclass[superscriptaddress,preprintnumbers,amsmath,amssymb,twocolumn,floats,aps,prl]{revtex4-2}
\usepackage{txfonts}
\usepackage{amssymb}
\usepackage{graphicx}
\usepackage{epstopdf}
\usepackage{sidecap}
\usepackage{CJK}
\usepackage{txfonts}
\usepackage{url}
\usepackage{color}
\usepackage{ulem}

\begin{document}

\title{Cu doping effects on the electronic structure of Fe$_{1-x}$Cu$_x$Se}

\author{S. Huh}
\thanks {Equal contributions}
\affiliation{Department of Physics and Astronomy, Seoul National University (SNU), Seoul 08826, Republic of Korea}
\affiliation{Center for Correlated Electron Systems, Institute for Basic Science (IBS), Seoul 08826, Republic of Korea}

\author{Z. Lu}
\thanks {Equal contributions}
\affiliation{Beijing National Laboratory for Condensed Matter Physics and Institute of Physics, Chinese Academy of Sciences, Beijing 100190, China}
\affiliation{School of Physical Sciences, University of Chinese Academy of Sciences, Beijing 100049, China}

\author{Y. S. Kim}
\affiliation{Department of Physics and Astronomy, Seoul National University (SNU), Seoul 08826, Republic of Korea}
\affiliation{Center for Correlated Electron Systems, Institute for Basic Science (IBS), Seoul 08826, Republic of Korea}

\author{D. Kim}
\affiliation{Department of Physics and Astronomy, Seoul National University (SNU), Seoul 08826, Republic of Korea}
\affiliation{Center for Correlated Electron Systems, Institute for Basic Science (IBS), Seoul 08826, Republic of Korea}

\author{S. B. Liu}
\affiliation{Beijing National Laboratory for Condensed Matter Physics and Institute of Physics, Chinese Academy of Sciences, Beijing 100190, China}
\affiliation{School of Physical Sciences, University of Chinese Academy of Sciences, Beijing 100049, China}

\author{M. W. Ma}
\affiliation{Beijing National Laboratory for Condensed Matter Physics and Institute of Physics, Chinese Academy of Sciences, Beijing 100190, China}
\affiliation{Songshan Lake Materials Laboratory, Dongguan, Guangdong 523808, China}

\author{L. Yu}
\email{li.yu@iphy.ac.cn}
\affiliation{Beijing National Laboratory for Condensed Matter Physics and Institute of Physics, Chinese Academy of Sciences, Beijing 100190, China}
\affiliation{School of Physical Sciences, University of Chinese Academy of Sciences, Beijing 100049, China}
\affiliation{Songshan Lake Materials Laboratory, Dongguan, Guangdong 523808, China}

\author{F. Zhou}
\affiliation{Beijing National Laboratory for Condensed Matter Physics and Institute of Physics, Chinese Academy of Sciences, Beijing 100190, China}
\affiliation{School of Physical Sciences, University of Chinese Academy of Sciences, Beijing 100049, China}
\affiliation{Songshan Lake Materials Laboratory, Dongguan, Guangdong 523808, China}

\author{X. L. Dong}
\affiliation{Beijing National Laboratory for Condensed Matter Physics and Institute of Physics, Chinese Academy of Sciences, Beijing 100190, China}
\affiliation{School of Physical Sciences, University of Chinese Academy of Sciences, Beijing 100049, China}
\affiliation{Songshan Lake Materials Laboratory, Dongguan, Guangdong 523808, China}

\author{C. Kim}
\email{changyoung@snu.ac.kr}
\affiliation{Department of Physics and Astronomy, Seoul National University (SNU), Seoul 08826, Republic of Korea}
\affiliation{Center for Correlated Electron Systems, Institute for Basic Science (IBS), Seoul 08826, Republic of Korea}

\author{Z. X. Zhao}
\affiliation{Beijing National Laboratory for Condensed Matter Physics and Institute of Physics, Chinese Academy of Sciences, Beijing 100190, China}
\affiliation{School of Physical Sciences, University of Chinese Academy of Sciences, Beijing 100049, China}
\affiliation{Songshan Lake Materials Laboratory, Dongguan, Guangdong 523808, China}

\date{\today}

\begin{abstract}
Using angle-resolved photoemission spectroscopy (ARPES), we studied the evolution of the electronic structure of Fe$_{1-x}$Cu$_x$Se from x = 0 to 0.10. We found that the Cu dopant introduces extra electron carriers. The hole bands near the $\Gamma$ point are observed to steadily shift downward with increasing doping and completely sink down below the Fermi level ($E_{F}$) for x $>$ 0.05. Meanwhile, the electron pocket near the M point becomes larger but loses the spectral weight near $E_{F}$. We also observed that effective mass of the electron band near the M point increases with doping. Our result explains why superconductivity disappears and metal insulator transition (MIT) like behavior occurs upon Cu doping in terms of electronic structure, and provide insight into emergent magnetic fluctuation in Fe$_{1-x}$Cu$_x$Se.
\end{abstract}
\maketitle

\section{I. Introduction}
Chemical substitution or doping studies have been widely conducted in correlated materials to control the electronic structure as well as physical properties. For example, it can change the chemical pressure, electron correlation, Fermi energy, carrier density. As such, investigating the chemical substitution effects can provide critical clues to understanding the underlying mechanism for various phenomena such as emergent magnetism, metal-insulator transition (MIT), spin/charge density wave, and superconductivity \cite{intromag1, intromit1, intromit2, introcdw1, introcdw2, introsdw1, introsc}.

Numerous chemical substitution studies have been conducted for iron based superconductors (IBSs) to understand the superconductivity mechanism and the interplay between the superconductivity and emergent phenomena \cite{introibs1, introibs2}. Among IBSs, chemical substitution studies have been intensively done on FeSe. The resultant changes of superconductivity and relevant electronic and magnetic properties through doping have provided rich insights of the underlining nature. Moreover, exotic phenomena such as Bardeen-Cooper-Schrieffer (BCS) Bose-Einstein condensation (BEC) crossover and topological superconductivity have been observed in substituted FeSe$_{1-x}$S$_{x}$ and FeSe$_{1-x}$Te$_{x}$ systems \cite{introbcs1, introbcs2, introtsc1, introtsc2}. While these studies expanded the boundary of the physics in FeSe, most of the studies have been limited to substitution of Se with isovalent atoms. There are relatively less doping studies on the Fe site due to the challenge in synthesizing high quality single crystals.

Investigating the Fe-site substitution may provide new opportunities for a better understanding of superconductivity and exploration of emergent phenomena. Cu doped FeSe is a suitable candidate to perform such studies, because of the recent success of high quality single crystal synthesis \cite{introgrowth1}. Previous studies on Cu doped FeSe reported that the structural transition and superconductivity are suppressed by a small amount of Cu doping and that a metal-insulator transition (MIT) like behavior appears with further doping \cite{introgrowth1, introgrowth2}. This unexpected MIT in FeSe call for systematic spectroscopic studies on the Cu doping eﬀect \cite{introcfese}. In addition, emergence of magnetic fluctuation \cite{intromag} and restoration of T$_{C}$  near 30 K under hydrostatic pressure \cite{intropressure} make this Fe$_{1-x}$Cu$_x$Se even more interesting and require deeper understanding of its properties in microscopic level. In elucidating the Cu doping eﬀect and mechanism of new emergent phenomena, electronic structure information should be vital, which have yet been lacking.

In this paper, we report the angle-resolved photoemission spectroscopy (ARPES) measurements of the electronic structure of Fe$_{1-x}$Cu$_x$Se (x = 0, 0.02, 0.03, 0.05, and 0.10). We ﬁnd that the Fermi surface is strongly Cu doping dependent. The size of the hole (electron) pocket near the $\Gamma$ (M) point decreases (increases) with doping. Upon the doping increased to x $>$ 0.05, the hole bands near  $\Gamma$ point sink down below the Fermi level ($E_{F}$), which changes the overall Fermi surface topology. In addition, we observe the spectral weight is reduced near $E_{F}$ as well as an increased effective mass of the electron band near the M point. Our observation reveals for the first time the electronic structure evolution of Fe$_{1-x}$Cu$_x$Se system in detail. It attributes the MIT behavior to electronic structure instead of merely impurity scattering. Our results may provide microscopic experimental foundation for understanding the emergent magnetic ﬂuctuation in this doping range.

\section{II. Experiment}

High quality single crystals of Fe$_{1-x}$Cu$_x$Se were synthesized by the chemical vapor transport method using mixture of AlCl$_{3}$ and KCl as transport agent \cite{introgrowth1}. In-plane electrical transport measurements were performed with a Quantum Design PPMS-9 system. ARPES measurements were performed using a home lab based system at Seoul National University. All spectra were acquired using a VG-Scienta electron analyzer and a discharge lamp from Fermi instruments. 21.218 eV photon energy was used in the experiment. Instrumental energy resolution was better than 20 meV. Samples were cleaved in an ultrahigh vacuum better than $8 \times 10^{-11}$ Torr. To minimize the aging effect, all data were taken within 10 hours after the cleaving.

\section{III. Results and discussion}

\begin{figure}[!h]
\includegraphics[width=8.6cm]{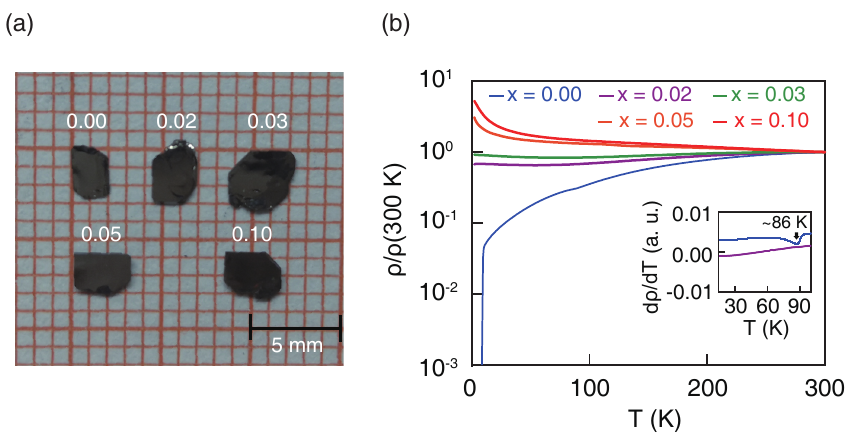} \caption{(Color online) (a) Optical image of Fe$_{1-x}$Cu$_{x}$Se single crystals. (b) Temperature dependent resistivity between 2 and 300 K. The inset shows the derivative of the resistivity (x = 0 and 0.02) as a function of temperature. The black arrow indicates structural transition (T$_{S}$).}
\end{figure}

Figure 1(a) shows an optical image of Fe$_{1-x}$Cu$_{x}$Se single crystals. Sizable single crystals with various doping concentration used in our ARPES experiments were obtained. The chemical composition of crystals was determined by inductively coupled plasma atomic emission spectroscopy. Temperature dependent in-plane resistivity for various doping is shown in Figure 1(b). Consistent with previous results \cite{introgrowth1, introgrowth2}, T$_{C}$ and structural transition temperature (T$_{S}$) are completely suppressed when Cu is only slightly doped (x $<$ 0.02). For a better comparison, we plot the first derivative of the resistivity in the inset. There is an anomaly near 86 K for x = 0 sample which indicates T$_{S}$, while no signature of such anomaly is observable at x = 0.02. The heat capacity measurement gives a consistent T$_{S}$ value \cite{introgrowth1}. In addition to the suppression of the phase transition, MIT like behavior is observed from the doping dependence of resistivity data. An upturn in the resistivity curve was observed for x = 0.05 and 0.10 while metallic behavior is observed over the entire temperature range for lower doping concentrations. Similar MIT behavior is reported for Cu doped iron pnictide systems \cite{cnares1, cnares2}, which implies that the Cu doping effect may be universal in IBSs.

We performed ARPES experiments to investigate the Cu doping effect on the electronic structure of Fe$_{1-x}$Cu$_{x}$Se. Figures 2(a)-(e) show Fermi surface maps of Fe$_{1-x}$Cu$_{x}$Se for different Cu concentrations. FeSe data were taken at 100 K to avoid the change in the band structure from the nematic phase transition which occurs near 90 K \cite{ts}, while other data were taken at 20 K since no structural transition is observed for these concentrations as mentioned above. This allows us to systematically compare the band structures of Fe$_{1-x}$Cu$_{x}$Se with various x values. With increasing Cu doping, the hole pocket near $\Gamma$ gradually decreases and disappear eventually when x $>$ 0.05. On the other hand, the size of the electron pocket near $M_1$ and $M_2$ point increases. The overall change of the Fermi surface topology suggests that Cu substitution seems to systematically dope extra electron carrier into the FeSe system.

A number of studies reported that Cu doping to IBSs may result in hole doping, but it is certainly not the case for FeSe \cite{hole1,hole2}. In addition, the observed electron doping behavior is similar to that of Ba(Fe$_{1-x}$Cu$_{x}$)$_{2}$As$_{2}$ and NaFe$_{1-x}$Cu$_{x}$As \cite{cba, introcna1, introcna2}. It is noteworthy that the shape of the electron pocket becomes more circular in comparison to the two perpendicularly crossing elliptical shape Fermi surface pocket of pristine FeSe. Such big circular electron pockets in absence of hole pocket at the center suggest the interesting similarity of the FS topology for heavily electron doped FeSe systems such as intercalated FeSe \cite{245} and 1 monolayer FeSe grown on SrTiO$_3$ \cite{fese1ml}.
\indent
\onecolumngrid\
\begin{figure}[!h]
\includegraphics[width=17cm]{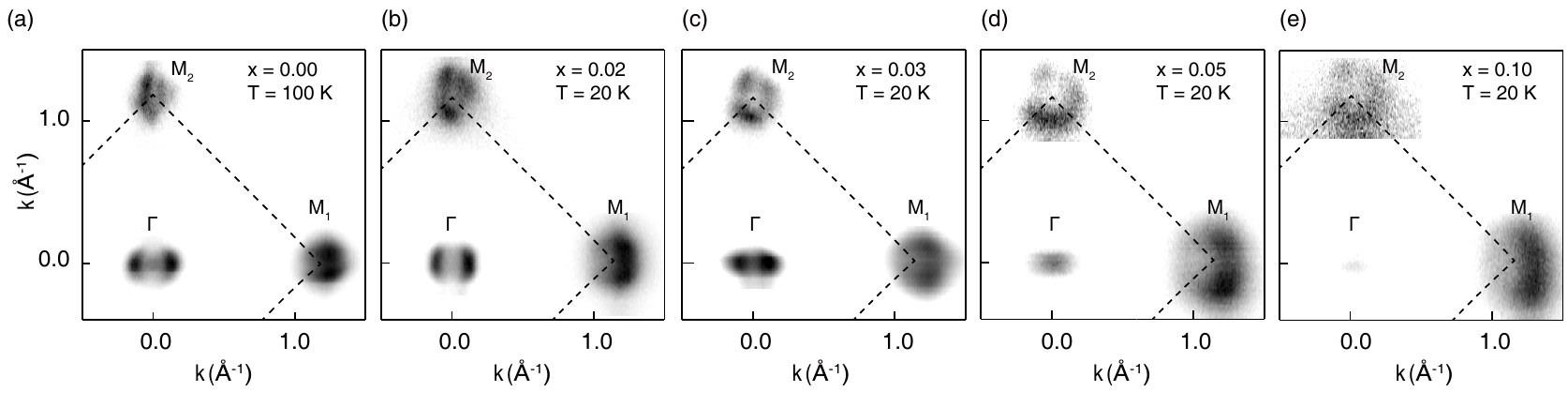}
 \centering \caption  {(Color online). (a)-(e) Fermi surface map of Fe$_{1-x}$Cu$_{x}$Se for different Cu concentrations. x = 0 data was taken at 100 K in order to avoid change in the band structure from the nematic phase transition, while others were taken at 20 K. All the data were taken by the same ARPES spectrometer with identical experimental setting parameters.}
\end{figure}
\twocolumngrid\
\indent

\indent

\begin{figure}[t]
\includegraphics[width=8.6cm]{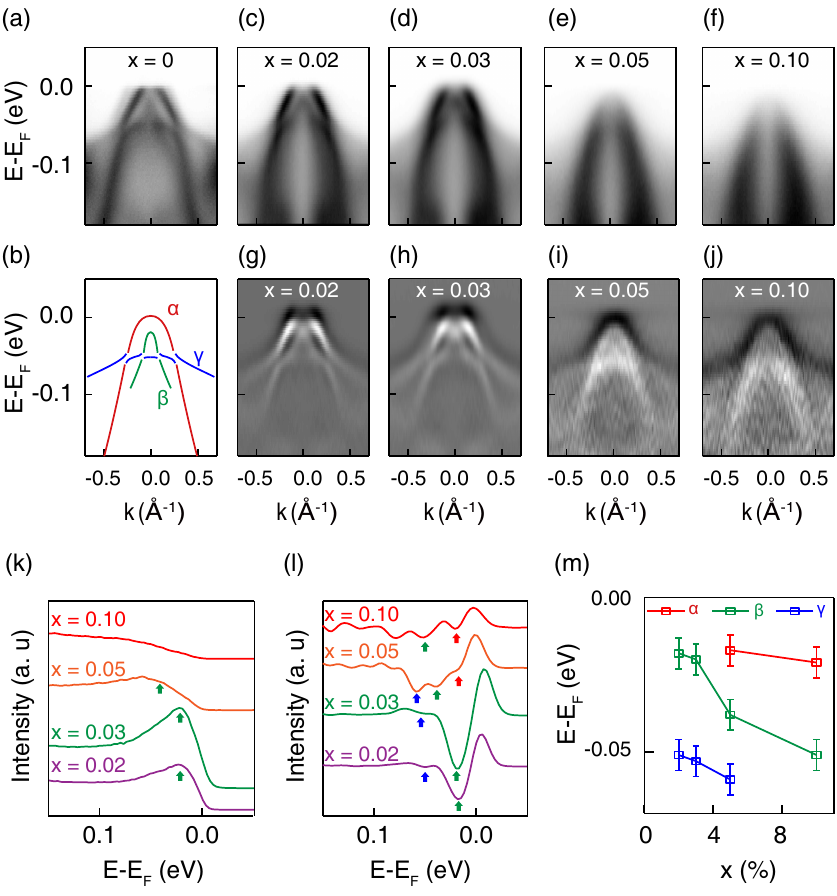} \caption{(Color online). (a) High symmetry cut and (b) schematic electronic structure of FeSe near the $\Gamma$ point. (c)-(f) High symmetry cut and (g)-(j) its second derivative data near the $\Gamma$ point for various x. Energy distribution curve (EDC) of (k) raw and (l) second derivative data. Red, green and blue arrows indicate the positions of the $\alpha$, $\beta$ and $\gamma$ bands, respectively. (m) Band positions as function of doping extracted from the EDCs plotted in panels (k) and (l).}
\end{figure}

In order to illustrate the evolution of electronic structure in details, we compare band dispersions along high symmetry directions. Figures 3(a) and (b) show a high symmetry cut of FeSe near the $\Gamma$ point and its schematic electronic structure, respectively. Similar to previous FeSe studies \cite{huh,liu,ishi}, so-called $\alpha$ and $\beta$ hole bands with $d_{xz/yz}$ orbital characters and $\gamma$ band with $d_{xy}$ orbital are well defined in our measurement. The effect of Cu doping can be resolved in this high symmetry spectra (Figures 3(c)-3(f)) and their corresponding second derivative spectra (Figures 3(g)-(j)). They clearly show that hole bands shift downward with doping. Especially, $\alpha$ hole band locates completely below $E_{F}$ from x $>$ 0.05, leaving strong reduction of spectral weight near the Fermi-level (Figure 3(k)), as a consequence of band gap opening. Such dramatic change of Fermi surface topology and spectral weight near $E_{F}$ can partially account for the insulating behavior appearing at doping x $>$ 0.05.Note that the weak residual hole pocket feature near the $\Gamma$ point in the Fermi surface maps of x = 0.05 and 0.10 in Figure 2 could come from broadening of dispersion tail possibly due to impurity scattering. In Figures 3(k) and (l), we plot energy distribution curves (EDC) of the raw data and their second derivative at the $k=0.0 {\text{\AA}}^{-1}$, respectively. Red, green and blue arrows indicate the topmost energy positions of the $\alpha$, $\beta$ and $\gamma$ bands, respectively. A summary on the band positions is shown in Figure 3(m), which clearly shows downward shift of hole bands through Cu doping. It needs to be emphasized that the doping effect seems more severe for $\alpha$ and $\beta$ bands compared to the $\gamma$ band. In other words, $d_{xz/yz}$ orbitals are more affected by doping than $d_{xy}$ orbital, implying an orbital selective doping dependence near the $\Gamma$ region.

\begin{figure}
\includegraphics[width=8.6cm]{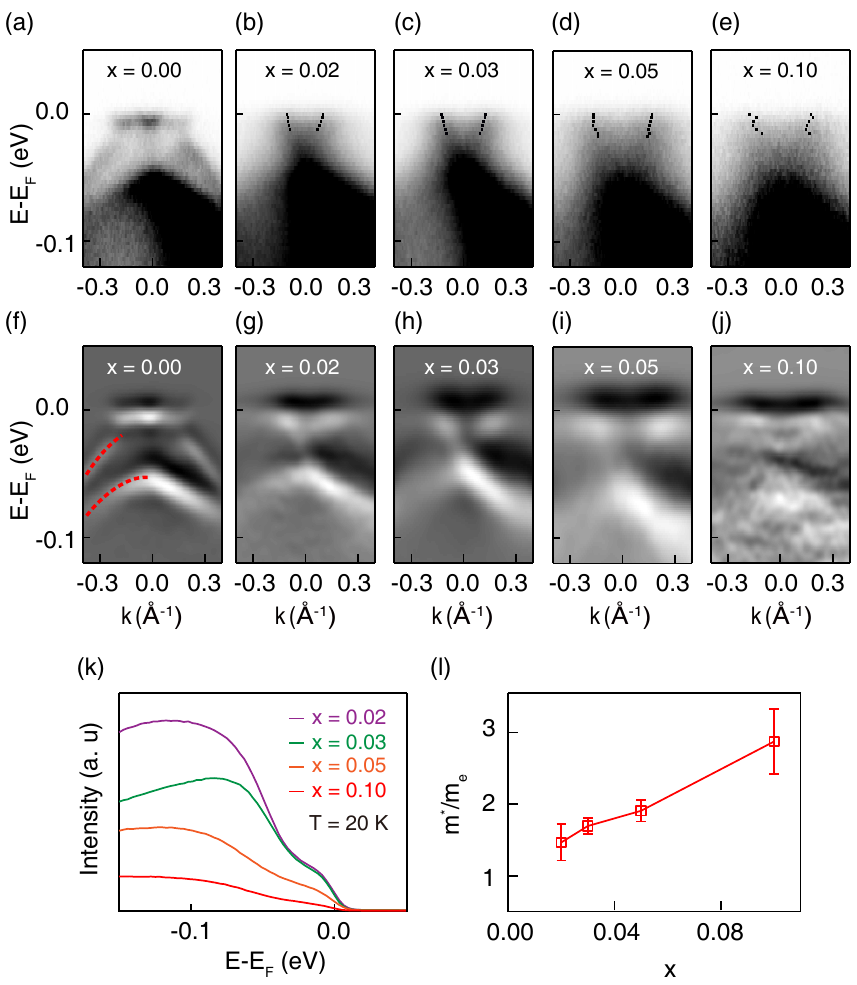} \caption{(Color online). (a)-(e) High symmetry spectra and (g)-(j) the corresponding second derivatives counterpart near the M$_2$ point. (k) Doping dependent angle integrated spectra near the M$_2$ point. Momentum integration window is ($-0.4 {\text{\AA}}^{-1}$, $0.4 {\text{\AA}}^{-1}$). (l) Effective mass of the electron band as a function of doping. $m^{*}$ is the mass of free electron.}
\end{figure}

Furthermore, the doping dependence of the band structures around the M point had also been examined. Figures 4(a)-(e) show high symmetry spectra data near the M$_{2}$ point measured at 20 K. As indicated in Figure 4(a), the splitting of the hole band (two red dash lines), as well known spectroscopic signature for the nematic phase, is not observed in Figures 4(b)-(e). This observation supports that nematic phase is suppressed upon Cu doping as mentioned above. For a better visualization, we also plotted the second derivative spectra in Figures 4(f)-(j). It is seen in the data that the electron pocket size increases with Cu doping. Here, only one electron band is resolved, because the low visibility of the $d_{xy}$ electron due to the matrix element effect \cite{huh, watson, cai}. Figure 4(k) shows doping dependent angle-integrated spectra over a momentum window of ($-0.4 {\text{\AA}}^{-1}$, $0.4 {\text{\AA}}^{-1}$). The spectral weight near $E_{F}$ gradually decreases as Cu is doped. In addition, eﬀective mass of electron band is determined by a parabolic ﬁt of the dispersion as shown in Figures 4(b)-(e). The effective mass almost doubly enhanced by Cu doping as shown in Figure 4(l). As mentioned above, the electron Fermi surface pocket near the M point looks similar to that of heavily electron doped FeSe systems which also show an enhanced effective mass compared to the pristine FeSe. Especially, the eﬀective mass for x = 0.10 is similar to that of 1 monolayer FeSe grown on SrTiO$_3$ \cite{mass}. Although Cu doped FeSe (x = 0.10) has a similar Fermi surface topology and correlation to 1 monolayer FeSe grown on SrTiO$_3$, the reason for its lack of superconductivity can be deduced from the impurity effect from Cu doping.

Nevertheless, our experimental observations suggest that the spectra weigh loss near $E_{F}$ around both $\Gamma$ and M points can account for the MIT in Cu doped FeSe, instead of simple impurity effect. Moreover, the enhanced correlation with Cu doping (larger effective mass) should also be considered as observed for the electron band at the M point. Emergence of a Mott insulating state in heavily Cu doped NaFeAs \cite{cnares1, mott} may also point to the importance of electron correlation in Cu doped FeSe system. Assuming that each Cu substitution results in doping of an electron, MIT appears when the total electron number of Fe 3d orbital is about 6.15 \cite{feseele}. Therefore, the standard Mott transition theory cannot account for insulating state of this system because the electron filling number is not an integer \cite{kwon}. Additional electronic/magnetic order or other factors from electron correlation should trigger the transition to an insulating phase.

Moreover, previous studies reported that magnetic fluctuation exists from x = 0.04 \cite{intromag}. On the other hand, our data show that the hole band near $\Gamma$ point is located below $E_{F}$ for a similar doping level. Thus, the origin of this magnetic fluctuation may not be solely explained within a Fermi surface nesting picture which is mostly believed to be responsible for the magnetism in IBS \cite{mazin}. Instead, changed Fermi surface topology and enhanced correlation should be certainly included to form theoretical model to explain the magnetism in Fe$_{1-x}$Cu$_{x}$Se. Concrete understanding requires further theoretical/experimental studies.

\section{IV. CONCLUSION}
We report results of a detailed electronic structure study on Fe$_{1-x}$Cu$_x$Se. We observe that the size of the hole (electron) pocket near the $\Gamma$ (M) point decreases (increases) with doping. With the increase of doping, we find that the hole band at the $\Gamma$ point is located below $E_{F}$, which changes the overall Fermi surface topology. In addition, it is found that the effective mass of the electron band at the M point increases upon Cu doping. Our observation explains, in terms of electronic structure, why the Fe$_{1-x}$Cu$_x$Se shows a MIT behavior and may provide clues to the mechanism of the magnetic fluctuation.

\section{Acknowledgment}
Authors would like to thank Prof. J. P. Hu, S. Ma, M. S. Kim, J. Y. Kwon and W. S. Kyung for helpful discussions. This work is supported by IBS-R009-G2 through the IBS Center for Correlated Electron Systems, and also supported by the National Natural Science Foundation of China (Grant Nos. 11834016, 11888101, 12061131005), the National Key Research and Development Projects of China (Grant No. 2017YFA0303003), the Key Research Program and Strategic Priority Research Program of Frontier Sciences of the Chinese Academy of Sciences (Grant Nos. QYZDY-SSW-SLH001, XDB33010200 and XDB25000000).

\end{document}